\documentclass[useAMS,usenatbib]{mn2e}

\def\var{\hbox{$\beta$ Lyrae}}

\def\dg{$^{\circ}$}

\usepackage{graphicx}
 \usepackage{times}
 \usepackage{lscape}

\title{On the accretion disc and evolutionary stage of $\beta$ Lyrae}
\author[ Mennickent \& Djura\u{s}evi\'{c}]
  {R.E. Mennickent$^{1}$\thanks{E-mail: rmennick@astro-udec.cl},
   G. Djura\v{s}evi\'c$^{2,3}$ \\
  $^1$Universidad de Concepci\'on, Departamento de Astronom\'{\i}a,
      Casilla 160-C, Concepci\'on, Chile\\
  $^{2}$ Astronomical Observatory, Volgina 7, 11060 Belgrade 38, Serbia   \\
 $^{3}$ Isaac Newton Institute of Chile, Yugoslavia Branch\\
  }
\date{}
 
\pagerange{\pageref{firstpage}--\pageref{lastpage}} \pubyear{2008}

\def\LaTeX{L\kern-.36em\raise.3ex\hbox{a}\kern-.15em
    T\kern-.1667em\lower.7ex\hbox{E}\kern-.125emX}

\begin{document}

\label{firstpage}

\maketitle

\begin{abstract}
We modeled the $V$-band light curve of $\beta$ Lyr  with two stellar components  plus an optically thick accretion disc around the gainer  assuming a semidetached configuration. We present the results of this calculation, giving physical parameters for the stars and the disc, along with general system dimensions.  We discuss the evolutionary stage of the system finding the best match with one of the  evolutionary models of Van Rensbergen et al.  According to this model, the system is found  at age 2.30 $\times$ 10$^{7}$ years, in the phase of rapid mass transfer, the second one in the life of this binary, in a Case-B mass-exchange stage with $\dot{M} = 1.58\times 10^{-5}$ M$_{\odot}$ yr$^{-1}$. This  result, along with
the reported rate of orbital period change and observational evidence of mass loss,   suggests
that the mass transfer in \var\, is quasi-conservative. The best model indicates that $\beta$ Lyrae finished a relatively large mass loss episode 31400 years ago.
The light curve model that best fit the observations has inclination angle $i$ = 81\dg, $M_{1}$ = 13.2 M$_{\odot}$,  $M_{2}$ = 3.0  M$_{\odot}$,  $R_{1}$ = 6.0 R$_{\odot}$ and  $R_{2}$ = 15.2 R$_{\odot}$.
The disc contributes 22\% to the total $V$-band light curve at quadrature, has a radius of 28.3 R$_{\odot}$ and the outer edge thickness is 11.2 R$_{\odot}$. The light curve model is significantly better with two bright regions in the disc rim with temperatures  10\% and 20\% higher than the disc outer edge temperature. We compare our results with earlier studies of this interacting binary.

\end{abstract}


\section{Introduction}

$\beta$ Lyrae  (Sheliak) is one of the best studied interacting binaries and also one of the more enigmatic bright objects in the sky. The literature for this object  is extensive and only
brief highlights will be given in this short introduction. The interested reader is directed to the reviews by Sahade (1980) and Harmanec (2002).
The current picture of the system  includes a B6-B8II donor star transferring mass onto an early B-type gainer. The hotter star is hidden by an optically and geometrically thick disc (Huang 1963; Wilson 1974; Hubeny \& Plavec 1991; Skulskii 1992, Linnell 2000) and some peculiar features can be explained by the presence of a bipolar jet emanating from the disc (Harmanec et al. 1996; Hoffman et al. 1998, Ak et al. 2007). 
Interferometry of $\beta$ Lyrae showed for the first time the direct image of a gravitationally distorted star  (Zhao et al. 2008; Schmitt et al. 2009). Non-conservative evolution was inferred from the presence of a radio nebula surrounding the binary system (Umana et al.  2000).
The system has an orbital period of 12.9 days increasing at a rate of 19 s yr$^{-1}$ (Harmanec \& Scholz 1993).
Among the notable properties of this system are: (1) the unexplained non-orbital 282.4 d photometric periodicity,  (2) the reversal of the depth of  primary and secondary eclipses
and their gradual evanescence  towards the ultraviolet and (3)
the presence of six systems of  infrared, optical and ultraviolet lines characterized by their morphology and radial velocity behavior (e.g. Harmanec 2002).

Many solutions for the system parameters have been proposed in the literature, some of them including an accretion disc around the gainer. In this paper we offer a complementary view of the system, and apply our light curve (LC)  synthesis code to fit the published $V$-band light curve,  obtaining a new set of representative system parameters. After obtaining these parameters, we search for a self-consistent evolutionary model by exploring a grid of theoretical evolutionary tracks for binary stars. The synthetic models are those by Van Rensbergen et al. (2008, 2011); they consider brief episodes of liberal  or non-conservative evolution. The paper is organized as follows.  In Section 2 we describe the light curve synthesis code including an accretion disc around the gainer and present the results of the light curve analysis. In Section 3 we explore the evolutionary stage of $\beta$ Lyr obtaining age and mass transfer rate and a theoretical set of system parameters. In Section 4 we provide a discussion of our results and a comparison with previous investigations. In Section 5 we give our conclusions.

\begin{table}
\caption{
Parameters of the best fit model for the {$\beta \ \rm Lyrae$} $V$-band light-curve
obtained by solving the inverse problem considering
an accretion disc around a  gainer in
critical rotation regime.}
 \label{}
      \[
        \begin{array}{llll}
            \hline
            \noalign{\smallskip}

{\rm Quantity} & & {\rm Quantity} & \\
            \noalign{\smallskip}
            \hline
            \noalign{\smallskip}
   n                               & 2852            & \cal M_{\rm_h} {[\cal M_{\odot}]} & 13.16\pm 0.3  \\
{\rm \Sigma(O-C)^2}                & 4.9040          & \cal M_{\rm_c} {[\cal M_{\odot}]} & 2.97 \pm 0.2  \\
{\rm \sigma_{rms}}                 & 0.0415          & \cal R_{\rm_h} {\rm [R_{\odot}]}  & 6.0  \pm 0.2  \\
   i {\rm [^{\circ}]}              & 86.1  \pm 0.5   & \cal R_{\rm_c} {\rm [R_{\odot}]}  & 15.2 \pm 0.2  \\
{\rm F_d}                          & 0.94  \pm 0.03  & {\rm log_{10}}g_{\rm_h}           & 4.0  \pm 0.1  \\
{\rm T_d} [{\rm K}]                & 8200  \pm 400   & {\rm log_{10}}g_{\rm_c}           & 2.5  \pm 0.1  \\
{\rm d_e} [a_{\rm orb}]            & 0.192 \pm 0.009 & M^{\rm h}_{\rm bol}               &-6.3  \pm 0.2  \\
{\rm d_c} [a_{\rm orb}]            & 0.01  \pm 0.01  & M^{\rm c}_{\rm bol}               &-4.7  \pm 0.1  \\
{\rm a_T}                          & 3.8   \pm 0.3   & a_{\rm orb}  {\rm [R_{\odot}]}    & 58.5 \pm 0.3  \\
{\rm f_h}                          & 20.2  \pm 0.5   & \cal{R}_{\rm d} {\rm [R_{\odot}]} & 28.3 \pm 0.3  \\
{\rm F_h}                          & 1.00            & \rm{d_e}  {\rm [R_{\odot}]}       & 11.2 \pm 0.2  \\
{\rm T_h} [{\rm K}]                & 30000 \pm 2000  & \rm{d_c}  {\rm [R_{\odot}]}       & 0.6  \pm 0.1  \\
{\rm A_{hs}=T_{hs}/T_d}            & 1.21  \pm 0.1   & \rm{log_{10} T_h}                 & 4.48          \\
{\rm \theta_{hs}}{\rm [^{\circ}]}  & 16.2  \pm 2.0   & \rm{log_{10} T_c}                 & 4.12          \\
{\rm \lambda_{hs}}{\rm [^{\circ}]} & 324.6 \pm 7.0   & \rm{log_{10} L_h}                 & 4.42          \\
{\rm \theta_{rad}}{\rm [^{\circ}]} & -13.5 \pm 5.0   & \rm{log_{10} L_c}                 & 3.81          \\
{\rm A_{bs}=T_{bs}/T_d}            & 1.12  \pm 0.1   &                                                   \\
{\rm \theta_{bs}} {\rm [^{\circ}]} & 53.8  \pm 5.0   &                                                   \\
{\rm \lambda_{bs}}{\rm [^{\circ}]} & 107.3 \pm 9.0   &                                                   \\
{\rm A_{h,c,d}}                    & 0.10, \ 0.35, \  0.59                                               \\
{\Omega_{\rm h}}                   & 12.12 \pm 0.02  &                                                   \\
{\Omega_{\rm c}}                   & 2.296 \pm 0.02  &                                                   \\
            \noalign{\smallskip}
            \hline
         \end{array}
      \]
Fixed parameters: $q={\cal M}_{\rm c}/{\cal M}_{\rm
h}=0.226$ - mass ratio of the components, ${\rm T_c=13300 K}$  -
temperature of the less-massive (cooler) donor, ${\rm F_c}=1.0$ -
filling factor for the critical Roche lobe of the donor,
${\rm F_h}=R_h/R_{zc}=1.0$ - filling factor for the
critical non-synchronous lobe of the hotter, more-massive gainer
(ratio of the stellar polar radius to the critical non-synchronous
lobe radius along z-axis for a star in critical rotation regime),
$f{\rm _c}=1.00$ - non-synchronous rotation coefficients
of the donor, ${\rm \beta_{h,c}=0.25}$
- gravity-darkening coefficients of the components.

\smallskip \noindent Note: $n$ - number of observations, ${\rm
\Sigma (O-C)^2}$ - final sum of squares of residuals between
observed (LCO) and synthetic (LCC) light-curves, ${\rm
\sigma_{rms}}$ - root-mean-square of the residuals, $i$ - orbit
inclination (in arc degrees), ${\rm F_d=R_d/R_{yc}}$ - disc
dimension factor (the ratio of the disc radius to the critical Roche
lobe radius along y-axis), ${\rm T_d}$ - disc-edge temperature,
$\rm{d_e}$, $\rm{d_c}$,  - disc thicknesses (at the edge and at
the center of the disc, respectively) in the units of the distance
between the components, $a_{\rm T}$ - disc temperature
distribution coefficient, $f{\rm _h}$ - non-synchronous rotation coefficient
of the more massive gainer (in the critical rotation regime),
${\rm T_h}$ - temperature of the gainer, ${\rm
A_{hs,bs}=T_{hs,bs}/T_d}$ - hot and bright spots' temperature
coefficients, ${\rm \theta_{hs,bs}}$ and ${\rm \lambda_{hs,bs}}$ -
spots' angular dimensions (radius) and longitudes (in arc degrees), ${\rm
\theta_{rad}}$ - angle between the line perpendicular to the local
disc edge surface and the direction of the hot-spot maximum
radiation, ${\rm A_{h,c,d}}$  - albedo coefficients of the system components and
the accretion disc, ${\Omega_{\rm h,c}}$ - dimensionless surface potentials
of the hotter gainer and cooler donor, $\cal M_{\rm_{h,c}} {[\cal
M_{\odot}]}$, $\cal R_{\rm_{h,c}} {\rm [R_{\odot}]}$ - stellar
masses and mean radii of stars in solar units,
${\rm log_{10}}g_{\rm_{h,c}}$ - logarithm (base 10) of the system components
effective gravity, $M^{\rm {h,c}}_{\rm bol}$ - absolute stellar
bolometric magnitudes, $a_{\rm orb}$ ${\rm [R_{\odot}]}$,
$\cal{R}_{\rm d} {\rm [R_{\odot}]}$, $\rm{d_e} {\rm [R_{\odot}]}$,
$\rm{d_c} {\rm [R_{\odot}]}$ - orbital semi-major axis, disc
radius and disc thicknesses at its edge and center, respectively,
given in solar units, $\rm{log_{10} T_{h,c}}$, $\rm{log_{10} L_{h,c}}$
- logarithm (base 10) of the system components effective temperature and
luminosity.
\end{table}

\section{Light curve analysis}

 \subsection{Model for an optically thick disc around the gainer}

Here we give a brief description of the  disc model that we apply to $\beta$ Lyrae.
The basic elements of the binary system model with a plane-parallel  disc and the corresponding light-curve synthesis procedure are described in detail by Djura\u{s}evi\'{c} (1992a, 1996). The model and code have been widely used, slightly improved  and tested
during our recent research of intermediate-mass interacting binaries (e.g. Djura{\v s}evic et al. 2010, 2011, 2012, Garrido et al. 2012, Mennickent et al. 2012).

We assume that the disc in $\beta$ Lyr is optically and geometrically thick. The disc edge is approximated by a cylindrical surface. In the current version of the model (Djura\u{s}evi\'{c}  et al. 2010), the  vertical thickness of the disc can change linearly with radial distance, allowing the disc to take a conical shape (convex, concave or plane-parallel), i.e. the disc presents
a triangular cross-section at both sides of the star. The geometrical properties of the disc are determined by its radius ($R_{d}$), its vertical thickness at the edge ($d_{e}$) and the vertical  thickness at the center or inner boundary  ($d_{c}$).

The cylindrical edge of the disc is characterized by its temperature, $T_{d}$, and the conical surface of the disc by a radial temperature profile obtained by modifying the temperature distribution proposed by Zola (1991):\\

$T (r) = T_{d} + (T_{h} - T_{d}) [1 - (\frac{r - R_{h}}{ R_{d} - R_{h}})]^{a_{T}}$ \hfill(1) \\

We assume that the disc is in physical and thermal contact with the gainer, so the inner radius and temperature of the disc are equal to the temperature and radius of the star ($R_{h}$, $T_{h}$). The temperature of the disc at the edge ($T_{d}$) and the temperature exponent ($a_{T}$ ), as well as the radii of the star ($R_{h}$) and of the disc ($R_{d}$) are free parameters, determined by solving the inverse problem (see Section 2.3).

The model of the system is refined by introducing active regions on the edge of the disc.
The existence of these active regions is supported by hydrodynamical simulations of \var\,
as we will discuss in Section 4.1.
The active regions have higher local temperatures so their inclusion results in a non-uniform
distribution of radiation. The model includes  two such active regions: a hot spot (hs) and a
bright spot (bs). These regions are characterized by their temperatures $T_{hs,bs}$ angular
dimensions (radius) $\theta_{hs, bs}$ and longitudes $\lambda_{hs, bs}$.
The longitude $\lambda$ is measured clockwise (as viewed from the direction of the +Z-axis,
which is orthogonal to the orbital plane) with respect to the line connecting the star centers (+X-axis),
in the range 0-360 degrees. 
These parameters are also determined by solving the inverse problem.

Due to the infall of an intensive gas-stream, the disc surface in the region of the hot spot becomes deformed as the material accumulates at the point of impact, producing a local deviation of radiation from the uniform azimuthal distribution.
In the model, this deviation is described by the angle $\theta_{rad}$  between the line perpendicular to the local disc edge surface and the direction of the hot spot maximum radiation  in
the orbital plane.

The second spot in the model, i.e. the bright spot,  approximates the spiral structure of an accretion disc, predicted by hydrodynamical calculations (Heemskerk 1994). The tidal force exerted by the donor star causes a spiral shock, producing one or two extended spiral arms in the outer part of the disc. The bright spot can also be interpreted as a region where the disc significantly deviates from the circular shape. For further information about  these active regions in the model,  see  Djura\u{s}evi\'{c}  et al. (2012).

One limitation of the code at its present implementation
is the lack of a detailed treatment for the donor irradiation by the disc part
facing the donor, including hot spot. This effect could be potentially
important in close binaries with a large difference in stellar and
disc temperatures. However, it should be a second-order effect compared
with the stellar and disc flux contributions (including donor irradiation by the gainer) already implemented in our code.
This is demonstrated in the very good fit  obtained with the orbital light curve.

Another limitation is the lack of consideration of a ``third light'', a source of continuum emission that is not eclipsed; in the case of $\beta$ Lyr it could be
the jet reported by Harmanec et al. (1996). 
 However, studies show that this third light has a minor influence in the total flux of \var, being larger in the ultraviolet, but much less evident in the optical (e.g. Linnell 2000). Hence, it is reasonable to assume that the optical flux is dominated by the stars and the disc.

\subsection{The $V$-band time series}

 We have used the $V$-band light curve and non-linear ephemeris of $\beta$ Lyrae published by Harmanec et al. (1996).
This dataset consists of 2582  $V$-band magnitudes obtained by differential photometry  in a 36 year interval at 12 different stations.
In most stations the reported mean $rms$ error per one observation of the corresponding check star is mostly between 0.01 and 0.03 mag.
The interested reader can find more details including the procedure of homogenization of the multi-station light curve in Harmanec et al. (1996).

Since the long photometric cycle of 282.4 d has a low amplitude of 0.024 mag, comparable to the error of an individual datapoint, viz.\,0.027, and this long cycle is probably not strictly periodic (Harmanec et al. 1996), we decided to work with the original $V$-band light curve without intending to remove the long cycle, which could introduce additional uncertainties. 


\subsection{The fitting procedure}

The light-curve fitting was performed using the inverse-problem solving method based on the simplex algorithm, and the model of a binary system with a disc described in the previous section. We used the Nelder-Mead simplex algorithm (see e.g. Press et al. 1992)
with optimizations described by Dennis and Torczon (1991).  While the direct problem comprises the calculation of the light curve from model parameters
given {\it a priori}, the inverse problem is the process of finding the set of parameters that
will optimally fit the synthetic light curve to the observations.  For more detail see e.g. Djurasevic (1992b).

To obtain reliable estimates of the system parameters, a good practice is to restrict the number of free parameters by fixing some of them to values obtained from independent sources.
Thus we fixed the spectroscopic mass ratio to $q$ = 0.226  and the donor temperature  to $T_{2}$ = 13300 $K$, based on published values  (e.g. Harmanec 2002, Linnell 2000). Our own LC simulations made with the parameters of the best solution but changing the mass ratio confirmed this $q$ value.
In addition, we set the gravity darkening coefficient and the albedo of the gainer and the donor to $\beta_{h,c}$  = 0.25 and $A_{h,c}$ = 1.0 in accordance with von Zeipel's law for radiative shells and complete re-radiation  (Von Zeipel 1924). The limb-darkening for the components was calculated in the way described by Djura\u{s}evi\'{c}  et al.  (2010).

The possible values of free
parameters are constrained by imposing the lowest and highest values  which seems reasonable based on previous studies of this binary. Here are the
ranges for the fitted parameters:

\begin{itemize}

\item Inclination: 70 to 90 degrees.
\item Disk dimension factor (the ratio of the disk radius and the radius of the critical Roche lobe
 along the y-axis): 0.5 to 1.0.
\item Disk edge temperature: 6000 to 12000 K.
\item Disk thickness: 0.01 to 0.3 distance between the components $a_{\rm orb}$.
\item The exponent of the disk temperature distribution: 1 to 5.
\item Non-synchronous rotation coefficient for the gainer (in critical rotation regime): 5 to 25.

\end{itemize}

After the first fit, these ranges are decreased according to the results of the first iteration.

We treated the rotation of the donor as synchronous,  i.e. we
used the non-synchronous rotation coefficient, defined as the
ratio between the actual and the Keplerian angular velocity,
$f_{c}$ = 1.0. Since it is assumed that the donor has filled its
Roche lobe (i.e. the filling factor of the donor was set to
$F_{c}$ = 1.0), the synchronization is expected to be the
consequence of the tidal effects. In the case of the gainer,
however, the accreted material from the disc is expected to
transfer enough angular momentum to increase the rate of the
gainer up to the critical velocity as soon as even a small
fraction of the mass has been transferred (Packet 1981, de Mink,
Pols \& Glebbeek 2007). This means that the gainer fills its
corresponding non-synchronous Roche lobe for  the star rotating in
the critical regime, and its dimensions and the amount of
rotational distortion are uniquely determined by the factor of
non-synchronous rotation. For $\beta$ Lyr we assumed critical
rotation for the gainer, and estimated a non-synchronous
rotation factor $f_h=20.2$ in the critical rotation regime. Our
additional modeling, obtained considering the synchronous rotation
regime for the gainer, does not differ much from the critical
rotation case.

We were able to model the asymmetry of the light curve very precisely by incorporating two regions of enhanced radiation on the  disc: the hot spot (hs), and the bright spot (bs).
The hot spot and the bright spot in our model are located on the edge side of the disk and
are described by the longitude of the center of the spot, the angular radius of the spot,
and the temperature ratio of the spot and  the unperturbed local temperature of the disk.
The difference between the temperature of the spot and the local unperturbed temperature
of the disk is what results in the difference in brightness. The hot spot can be located
at longitudes between 270 and 360 degrees, and the bright spot can be located at any
longitude. The angular radius of the spots was constrained to the range from 0 to 30 degrees
for the hot spot, and from 0 to 70 degrees for the bright spot; the temperature ratio for
the spots can be from 1.0 to 1.5.

\subsection{Results of the light curve fitting}





 Compared to the model without the active regions on the
disk, the model which includes the hot and bright spots (hs+bs)
fits the observations much better (see Fig.\,1).
The dominant feature of our optimal model is the bright spot,
and the parameters describing its size, temperature and
location are the most important for improving the fit. The
existence of a bright spot indicates significant deviation of
the disk from cylindrical shape, or other phenomena that are
not directly modeled, such as a spiral arm in the outer part of
the disk. The parameters of the hot spot don't have as much
influence on the quality of the fit because the bright spot is
significantly larger in size.

\begin{figure*}

\includegraphics[width=1.0\textwidth]{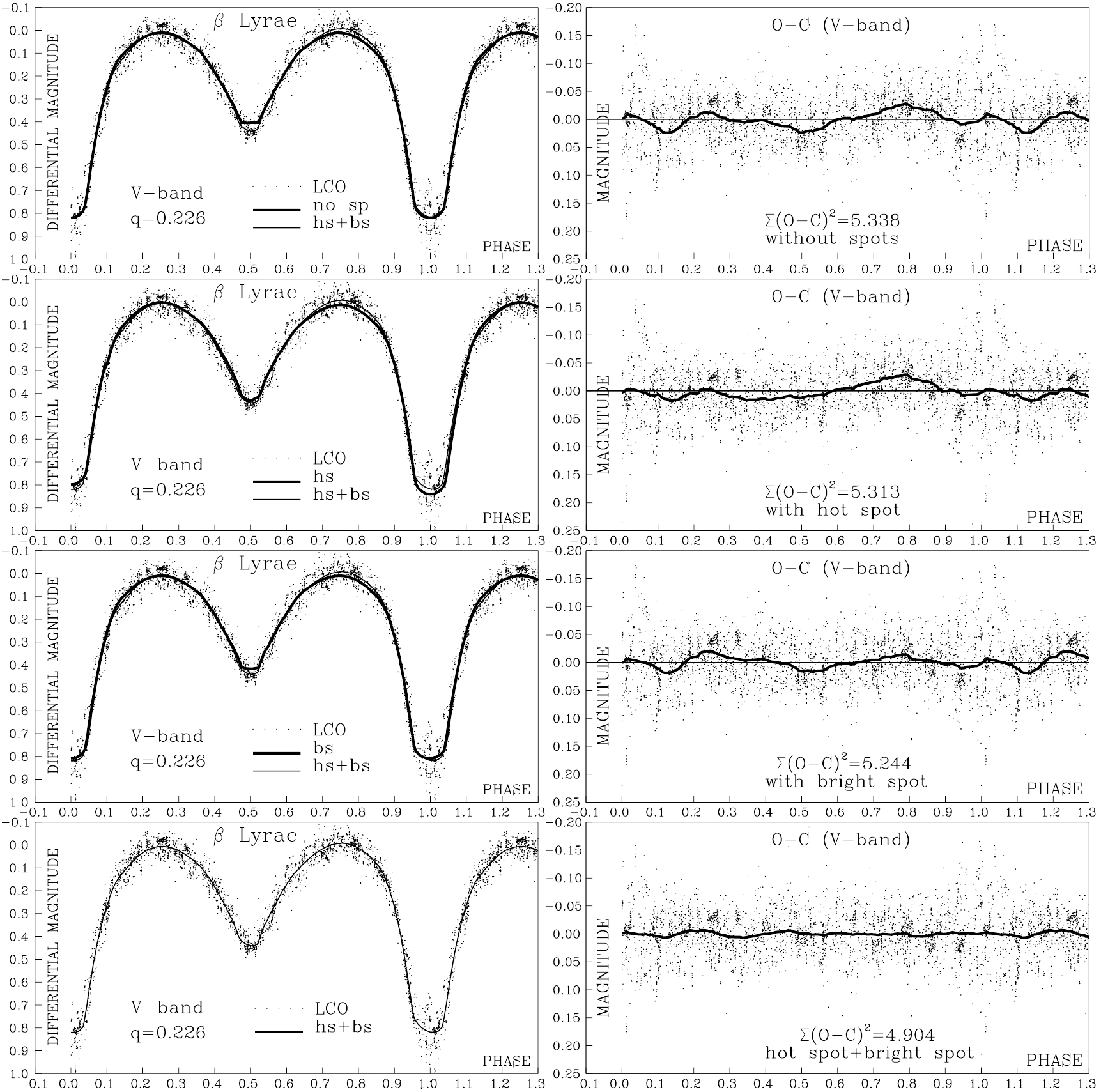}
\caption{Observed (LCO) and model's synthetic light-curves
(without spots - no sp, with hot spot - hs, with bright spot -
bs and with hot spot and bright spot - hs+bs on the accretion
disk edge) of {$\beta \ \rm Lyrae$} obtained by analyzing
photometric observations and the corresponding final O-C
residuals between the observed and model's synthetic light
curves. The O-C plots show also the final $\Sigma(O-C)^2$ for
the corresponding model and polynomial fit of the residuals
presented by the solid line. In the right-hand column a thin
horizontal line shows 0 magnitude.} \label{fbLyrae-s}
\end{figure*}

\begin{figure*}
\includegraphics[]{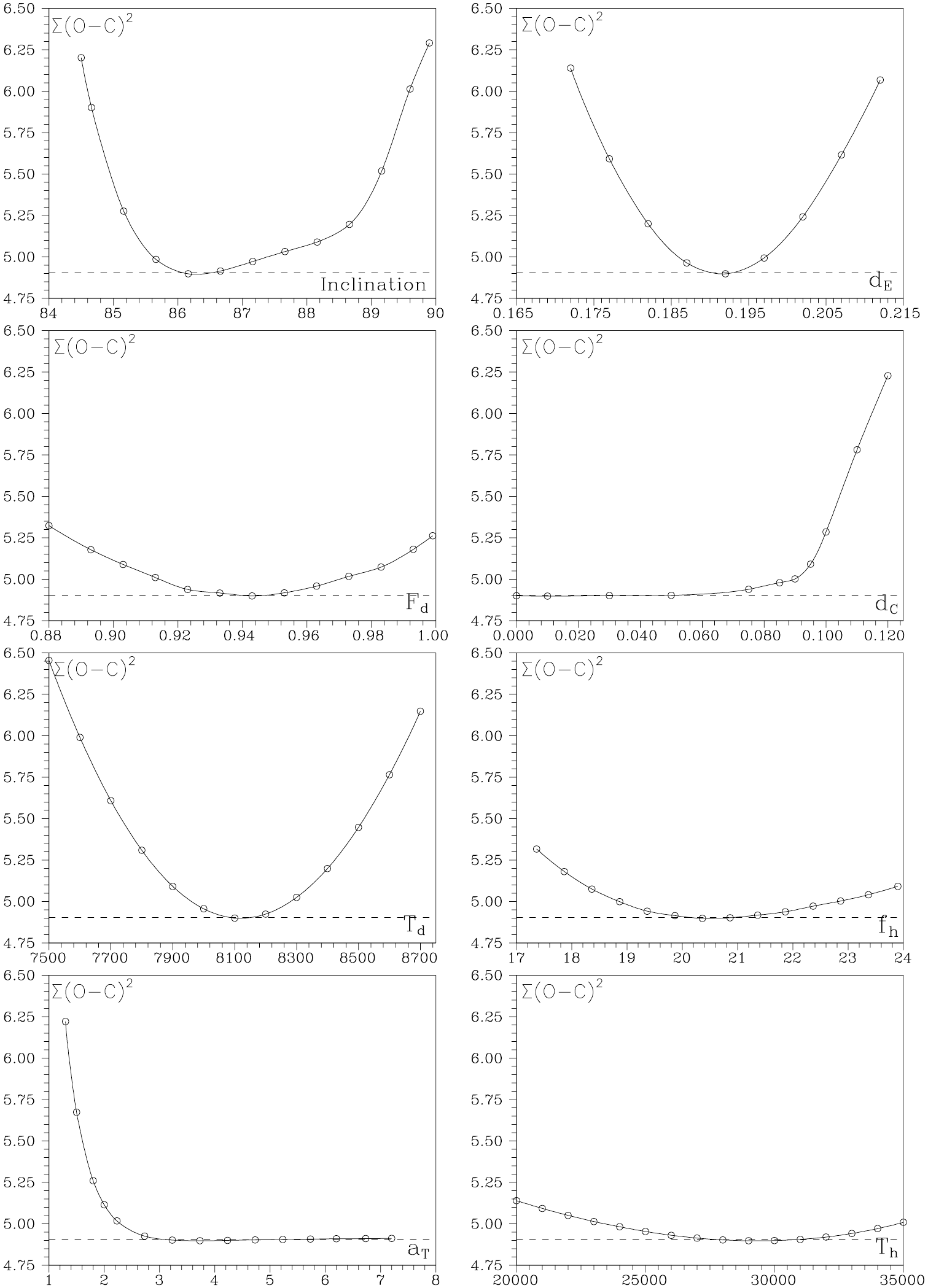}
\caption{ The dependence of $\Sigma(O-C)^2$ on basic model
parameters. Each panel shows the variation of one parameter
while all others are kept fixed on the optimal values obtained
for the best-fitting model, which includes the hot and bright
spots (hs+bs) on the accretion disk. The dashed line shows the minimum value of $\Sigma(O-C)^2 = 4.904$ in each
panel.} \label{sigma-par}
\end{figure*}


An attempt to model the system without including the active regions on the edge of
the accretion disk showed that the observed light curve cannot be described well with
such a model. This is especially noticeable in the secondary minimum and around the
secondary maximum where there is a significant asymmetry in the light curve (see Fig.\,1).
The O-C residuals between the observed (``LCO'') and synthetic light curve
(``no sp'') clearly show the deficiencies of this model for optimal fitting of the
observations.

If a hot spot region is added to the model of the disk, the model fits the observations
better in the secondary minimum, but still does not describe the asymmetry in the region
of the secondary maximum, and introduces problems with the fitting of the primary minimum.
The third model we tested has only a bright spot on the edge of the disk and gives a somewhat
better fit, but with significant differences between the model and the observations in
certain phase intervals (see Fig.\,1). Finally, Fig.\,1 shows the optimal model of the system,
which contains both the hot spot and the bright spot regions (Òhs+bsÓ) on the disk. The
picture clearly shows that this model can successfully fit the observations and the
light curve asymmetries that were problematic in other models we considered.


 Apart from this graphical presentation, the advantages of
our optimal model can be demonstrated by applying the Fisher
statistical F-test, which shows that the models without spots
($S_1=\Sigma(O-C)^2=5.338$) and the models with a single (hot -
$S_2=\Sigma(O-C)^2=5.313$ or bright -
$S_3=\Sigma(O-C)^2=5.244$) spot can be discarded with the
probability of 95\% when compared to the optimal model (hot
spot and bright spot - $S_{opt}=\Sigma(O-C)^2=4.904$). Namely,
the mentioned values give the following quantities:
$F_1=S_1/S_{opt}=1.0885$, $F_2=S_2/S_{opt}=1.0834$ and
$F_3=S_3/S_{opt}=1.0693$, while the Fisher statistical F-test
for the probability of 95\% gives the critical value of the
F-distribution F= 1.064. This clearly shows that the model with
active hot and bright regions on the disk fits the observations
much better than models without spots (F1) and with single hot
spot (F2) or bright spot (F3) because these $F_{1,2,3}$
calculated from the data are greater than the critical value of
the F-distribution for probability 95\%, as can also be seen
in Fig.\,1. On the other hand, adding more bright spot regions
does not improve the quality of the fit.

 Fig.\,2 shows how varying of basic model
parameters around the optimal values (obtained for the best
model, which includes the active regions on the disk) affects
the $\Sigma(O-C)^2$ value. Each panel shows the variation of
one model parameter while all others are kept fixed on their
optimal values. The scale in all plots is the same, to make 
easier to note which parameters influence the quality of the
fit at most.

The model is relatively insensitive to the temperature of the
gainer and to the exponent of the disk temperature
distribution. This is to be expected since the disk is seen
edge-on, and it covers a large part of the gainer, so the inner
regions of the disk don't contribute much to its total
radiation. The value found by us, viz. $T_{h}$ = 30.000 K, is
consistent with that used by other authors (e.g. Hubeny \&
Plavec 1991). The model also shows relatively low sensitivity
to changes in disk radius (parameter $F_{d}$), inner disk thickness
($d_{c}$), and the non-synchronous rotation coefficient of the
gainer ($f_{h}$). It is more sensitive to changed in the orbital
inclination, the outer disk thickness ($d_{e}$) and temperature
($T_{d}$).



\begin{figure}
\includegraphics[width=0.42\textwidth]{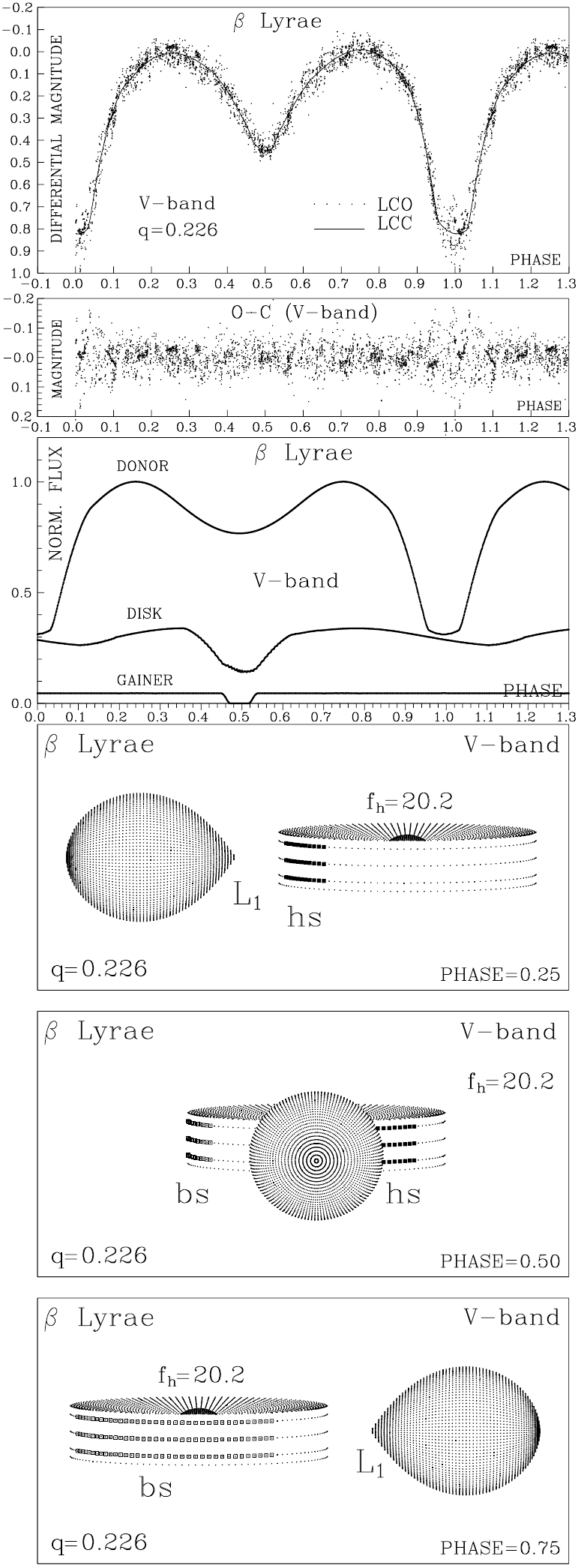}
\caption{Observed (LCO) and synthetic (LCC) light curves of
$\beta$ Lyr, obtained by analyzing photometric observations;
final O-C residuals between the observed and optimum synthetic
light curves; fluxes of the donor, the gainer and of the disc,
normalized to the donor flux at phase 0.25; the views of the
optimal model at orbital phases 0.25, 0.50 and 0.75, obtained
with parameters estimated from light curve analysis. We
indicate the mass ratio ($q$), the approximate positions of the
inner Lagrangian point ($L_{1}$), the hot spot ($hs$), the
bright spot ($bs$), and the gainer non-synchronous rotation
factor defined in the text ($f_{h}$).}
\label{lcs}
\end{figure}

\subsection{The best light-curve model}

The  fit, $O-C$ residuals, individual donor, disc and gainer flux contributions and the view of the optimal model at orbital phases 0.25, 0.50 and 0.75, obtained with the parameters estimated by the light curve analysis, are illustrated in Fig.\,3. We note that residuals show no dependence on orbital or long-cycle phases, except a larger random scatter around main eclipse, a feature already evident in the light curve.
 We find that the best fit model of \var\,contains an optically and geometrically thick  disc around the hotter, more massive gainer star (Table 1). With a radius of $R_{d} \approx 28.3 R_{\odot}$, the disc is 4.7 times larger than the central star ($R_{h} \approx 6.0 R_{\odot}$). The disc has a convex shape, with central thickness $d_{c} \approx 0.6R_{\odot}$ and edge thickness $d_{e} \approx 11.2 R_{\odot}$. The temperature of the disc increases from $T_{d}$ = 8200 $K$ at its edge, to $T_{h}$ = 30000 $K$ at the inner radius, where it is in thermal and physical contact with the gainer. The relatively large disc temperature gradient explains the big difference between disc thickness at the inner and outer edges.

In the best model the hot spot with 16\dg angular radius covers 9\% of the disc outer rim and it is
situated at longitude $\lambda_{hs} \approx$  325\dg, roughly between the components of the system,
at the place where the gas stream falls onto the disc (Lubow \& Shu 1975). 
The temperature of the hot spot is approximately 21\% higher than the disc edge
temperature, i.e. $T_{hs} \approx 9900$ $K$.
Although including the hot spot region into the model improves the
fit, it cannot explain the light- curve asymmetry completely. By
introducing one additional bright spot, larger than the hot spot
and located on the disc edge at $\lambda_{bs} \approx$ 107\dg, the
fit becomes much better. This bright spot has $T_{bs} \approx$
9200 $K$ and with 54\dg angular radius covers 30\% of the
disc outer rim.

A Fourier analysis (Lenz \& Breger 2005) and a Phase Dispersion Minimization analysis (Stellingwerf 1978) of residuals of the best fit model
confirm the long periodicity of 282 days found by Harmanec et al. (1996) with the same dataset but with other light curve model.



\section{On the evolutionary stage of $\beta$ Lyr}

In this section we compare the system parameters obtained in Section 2.2 and listed in Table 1 with those predicted by binary evolution models including epochs of non-conservative evolution. We inspected the 561 conservative and non-conservative evolutionary tracks  by Van Rensbergen et
al.  (2008) available at the Center de Donn\'ees Stellaires (CDS) looking for the best match for the system parameters found for $\beta$ Lyr.
Models with strong and weak tidal interaction were studied, although only the  latter ones should allow critical rotation of the gainer.
Following Mennickent et al. (2012), a  multi-parametric fit was made with the  synthetic ($S_{i,j,k}$)  and observed ($O_{k}$) stellar parameters  mass, temperature, luminosity and radii (obtained from the light-curve fitting), and the orbital period,  where $i$ (from 1 to 561) indicates the synthetic  model, $j$ the time $t_{j}$ and $k$  (from 1 to 9)  the  stellar or orbital parameter.
 Non-adjusted parameters were mass loss rate, Roche lobe radii, chemical composition, fraction of accreted mass lost by the system and age.
 For every synthetic model $i$ we calculated the quantity $\chi^{2}_{i,j}$ at every $t_{j}$ defined by:\\

$\chi^{2}_{i,j} \equiv (1/N) \Sigma_{k} w_{k}[(S_{i,j,k}-O_{k})/O_{k}]^{2} $\hfill(2) \\

\noindent
 where $N$ is the number of observations (9) and $w_{k}$ the statistical weight of the parameter $O_{k}$, calculated as:\\

$w_{k} = \sqrt{O_{k}/\epsilon(O_{k})} $\hfill(3) \\

\noindent
where $\epsilon(O_{k})$ is the error associated to the observable $O_{k}$.
The model with the minimum $\chi^{2}$ corresponds to the model with the best evolutionary history of $\beta$ Lyr. The absolute minimum $\chi^{2}_{min}$  identifies the age of the system along with the theoretical stellar and orbital parameters. The high accuracy of the orbital period dominates the search for the best solution in a single evolutionary track (Fig.\, 5), but the others parameters play a role when comparing tracks corresponding to different initial stellar masses.

We find the absolute $\chi^{2}$ minimum  (viz.\,0.0291) in the weak
  interaction model with initial masses of 12 and 7.2 M$_{\odot}$ and initial orbital period of 2.5  days; their parameters are shown in Table 2.
  All other models give larger $\chi^{2}$ and were rejected as reliable solutions. A comparison  of  $\chi^{2}$ for these models
  with  the optimal solution indicates that the best fit  stands out among other solutions  in the multi-parametric space (Fig.\,6). The solution very near to the absolute minima in Fig.\,6 (easily visible in the bottom panels) corresponds to the strong interaction model for the same initial conditions; it has $\chi^{2}$ = 0.0308 and  their parameters, also
  given in Table 2 for reference, differ 
by a  small and almost always neglectable amount from the adopted parameters.

The corresponding evolutionary tracks for the gainer and
donor stars are shown in Fig.\,4, along with the position
for the best model for $\beta$ Lyr. We observe a relatively good
match for the gainer star parameters but a small mismatch with the
donor temperature and luminosity. 
The observed donor is slightly
fainter and cooler than the model prediction. The difference
amounts to 0.9\% in $log\,T$ and 2.7\% in $log\,L$. 
 This small discrepancy remains when considering the next more reliable solution, those
with strong tidal interaction and whose parameters are given in Table 2. 
We propose that this 
mismatch could be due to a limitation of the existing grid of
synthetic models available, that are not dense enough in initial
orbital period and initial masses to give  account of all possible
evolutionary combinations, and eventually could not reproduce the
\var\, evolution perfectly. For instance, only initial masses for
the donor of 7, 8, 9, 12 and 15   M$_{\odot}$ are considered in
the range compatible with \var. Whereas this discretization of
initial parameters probably does not produce a large discrepancy
with the actual system parameters (as suggested by the absence of competent solutions near  the absolute minima
in Fig.\,6), it is likely that small differences between observed
and synthetic parameters, as observed with the donor, occur. At present, we cannot quantify this effect.

The best fit  indicates that  $\beta$ Lyr is found inside a burst of mass transfer, the second one in the life of this binary (Fig.\,5). The donor is an inflated ($R_{c}$ = 14.4 R$_{\odot}$) and evolved 2.2 M$_{\odot}$ star with its core completely exhausted of hydrogen. According to the best model the system has now an age of 2.3 $\times$ 10$^{7}$ yr and
$\dot{M}$ =  1.58 $\times$ 10$^{-5}$  M$_{\odot}$ yr$^{-1}$. The system reverted its mass ratio not at the present
semidetached stage, but about 8 Myr ago, during a much stronger episode of mass transfer.
The small difference between $\dot{M_h}$ and  $\dot{M_c}$ in Table 2 is a purely numeric effect due to the iterative process of calculation; the transferred mass equals the accreted mass in the long-term, i.e. according to the best model $\beta$ Lyrae is in a conservative mass transfer stage. An inspection of the best-model evolutionary track indicates that:   (i) the system is found 147.000 years after 2nd Roche lobe contact
and (ii) in this interaction episode the system has lost (previously) 0.03 M$_{\odot}$ into the interstellar medium and the gainer has eventually accreted  1.55  M$_{\odot}$. The last figures were derived from the mass balance among two epochs in the evolutionary track: the epoch corresponding to the start of mass loss and the present epoch.
However,
if critical rotation impedes the accretion of more mass (something not considered by the binary evolution codes), part of this mass
could have been ejected into the interstellar medium or remains still stored  in the accretion disc.

\begin{figure*}
\scalebox{1}[1]{\includegraphics[angle=0,width=17 cm]{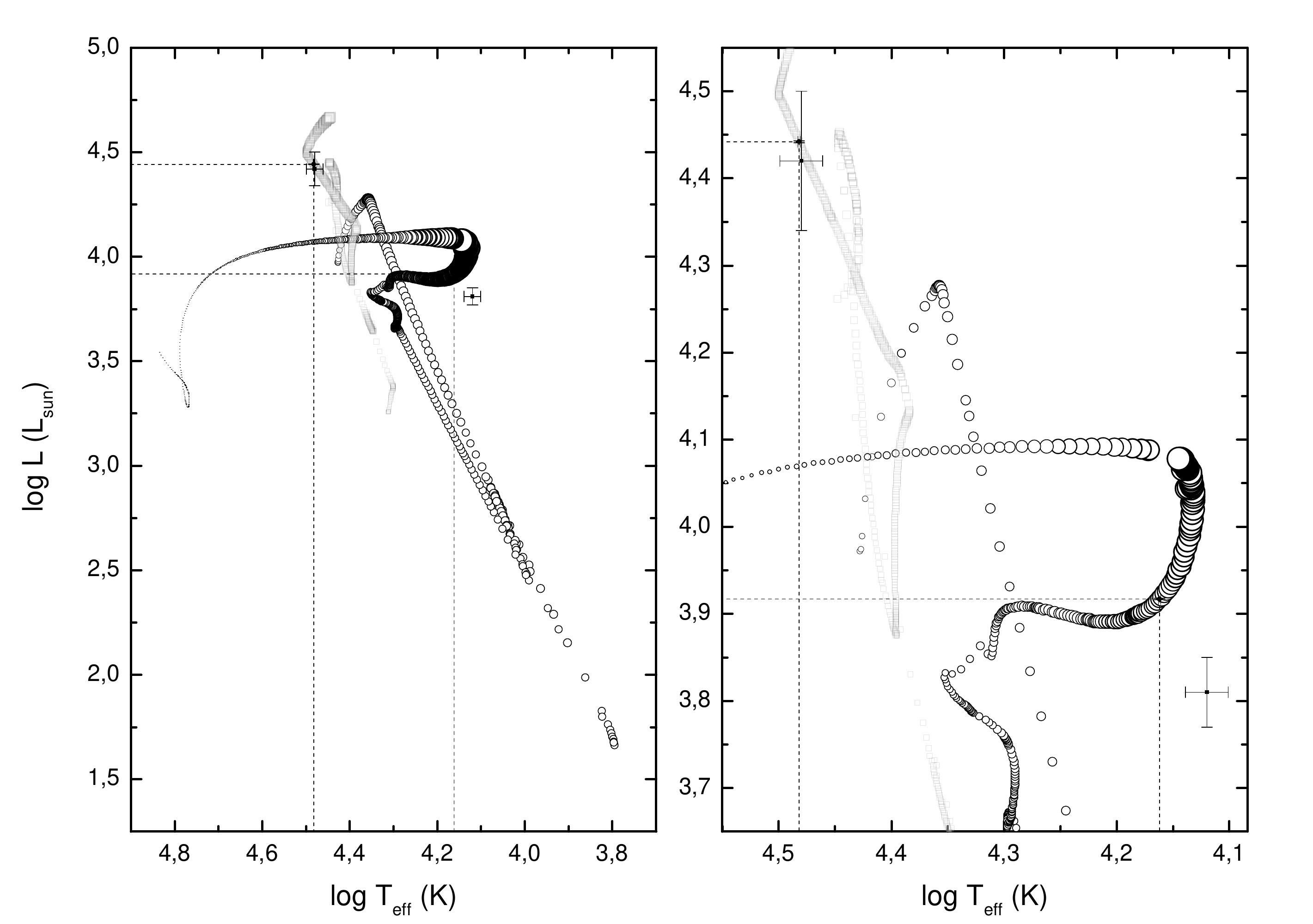}}
\caption{Evolutionary tracks for the  binary star model from Rensbergen et al. (2008) that best fit the data.
Donor (right black track) and gainer (left gray track) evolutionary paths are shown,
along with the  parameters derived from the LC fit.
The best fit is reached at the time corresponding to the model attached to the axis by dashed lines, that is  characterized in Table 2.
The mismatch for the donor is discussed in the text. Stellar
sizes are proportional to the circle diameters. An expanded view is shown in the right panel where half of the points
are shown for the donor track  for best visualization.
}
  \label{x}
\end{figure*}

\begin{figure}
\scalebox{1}[1]{\includegraphics[angle=0,width=8.5cm]{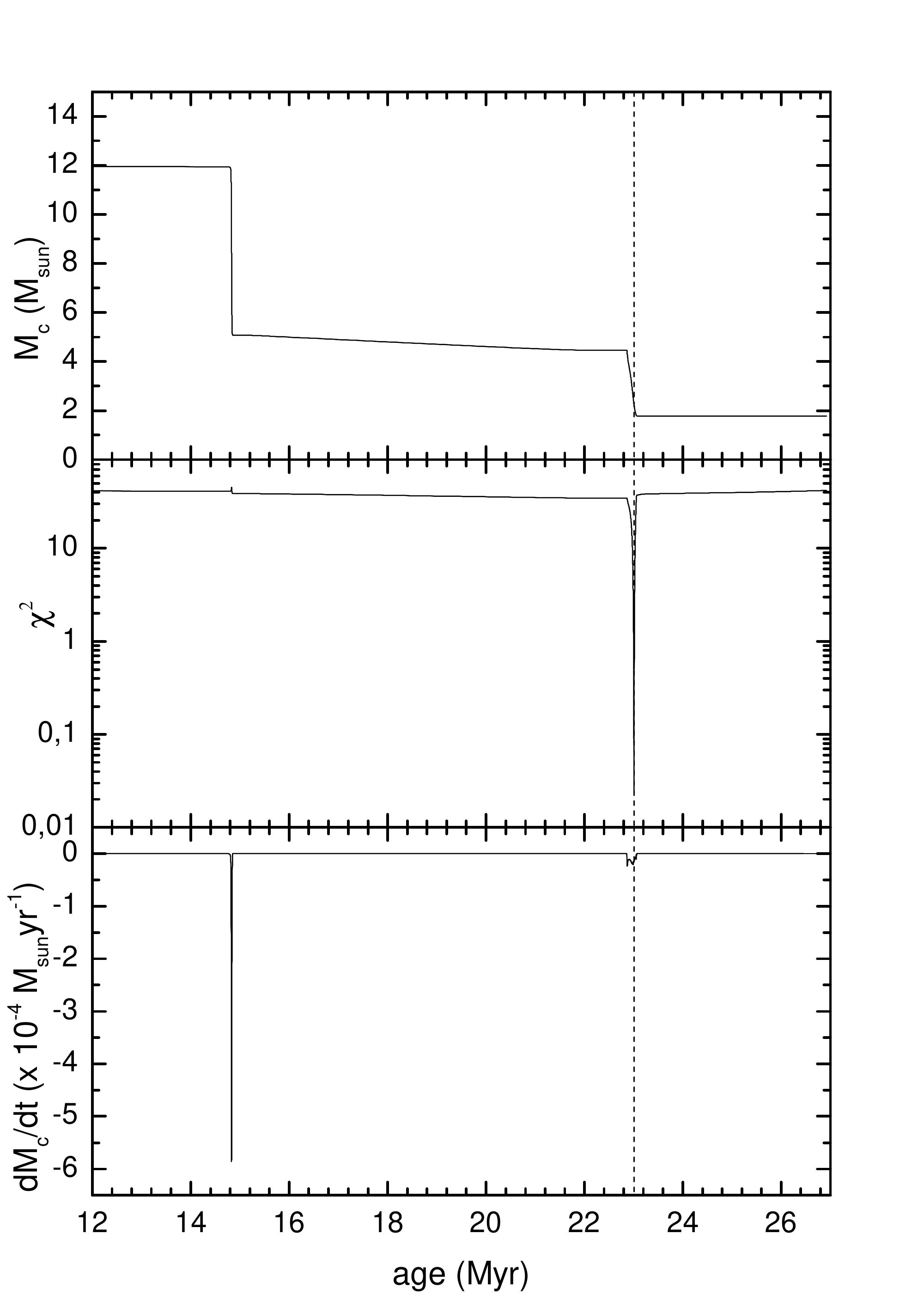}}
\caption{
  $M_{c}$,  $\chi^{2}$ (defined in Eq.\,2)  and $\dot{M_c}$ for the best evolutionary track. The vertical dashed line  indicates the position for the best model.}
  \label{x}
\end{figure}

\section{Discussion}

In this section we compare our \var\, parameters with those previously reported in the literature. We also give a semblance of the future appearance of the system, just after
finishing the mass transfer process, based on the prediction of the best synthetic model.

\subsection{Comparison with published system and disc parameters}

A range of stellar masses, inclinations and mass ratios have been
published for \var\,during the past 50 years, but the latest calculations seems to indicate  gainer and  donor masses around 13 $M_{\odot}$ and  3 $M_{\odot}$, respectively (Harmanec 2002).
Our stellar masses, listed in Table 1, are consistent with these numbers.




Similarly to other authors (e.g. Hubeny \& Plavec 1991, hereafter HP, and Linnell 2000), we have found that a geometrically thick disc significantly improves the fit to the light curve in \var.
In our model the disc contributes 22\% to the total $V$-band light curve at quadrature,
 has a radius of 28.3 R$_{\odot}$ (HP find 25.5 R$_{\odot}$ and Linnell 30 R$_{\odot}$) and the thickness goes from 0.58 R$_{\odot}$ at the inner edge to 11.2 R$_{\odot}$ at the
rim  (HP find 0.34 R$_{\odot}$  to 14 R$_{\odot}$,  respectively, and Linnell finds 16 R$_{\odot}$ at the rim).

Our derived mass transfer rate $\dot{M} = 1.58\times 10^{-5}$ M$_{\odot}$ yr$^{-1}$ matches the canonical value commonly used in numerical simulations (1-4 $\times$ 10$^{-5}$ M$_{\odot}$ yr$^{-1}$, e.g. Bisikalo et. al. 2000, Nazarenko \& Glazunova 2003). This canonical value is often calculated by assuming conservative mass transfer and the observed value for the orbital period change (19 s yr$^{-1}$;  Harmanec \& Scholz 1993). In fact,  a binary interchanging material in  a conservative way has a constant period change given by:\\

$\frac{\dot{P}}{P} = -\frac{3(1-q) \dot{M_{2}}}{M_{2}}$ \hfill(4) \\

\noindent (Huang 1963). For $\beta$ Lyr this equation gives $\dot{M}$ = (2.2 $\pm$ 0.2) $\times$ 10$^{-5}$ M$_{\odot}$ yr$^{-1}$, where the major contribution
to the error budget comes from the donor mass uncertainty quoted in Table 1.

We notice two important issues here: (1) The best fit model is conservative, but observations (discussed in the next paragraph) suggest mass loss  at the present epoch and (2) our $\dot{M}$ value is near the conservative figure derived from Eq.\,4. Whereas the first point could indicate that the models still do not reproduce the actual physics of mass loss processes,
the second point suggests
 that the mass loss rate probably is much smaller than the mass transfer rate,  producing a mass transfer practically conservative or, in other words, quasi-conservative.

Jets and nebular radio emission have been considered signatures of mass loss in \var. From the evolutionary track discussed in Section 3, we have estimated that 0.03 M$_{\odot}$ have been lost in the present episode of Roche lobe overflow. Umana et al. (2000) estimate 0.015 M$_{\odot}$ from the analysis of the radio nebula but using $\dot{M}$  $\sim$ 10$^{-7}$ $M_{\odot}$ yr$^{-1}$ for the donor wind ionized by the radiation field of the hotter companion.
An earlier estimate  based on
the mass exchange models by De
Greve and Linnell (1994, hereafter DGL94) is 0.43 M$_{\odot}$.  Our mass loss estimate is linked to the nature of the mass loss process considered in the Van Rensbergen et al. (2008, 2011) models. These models consider mass loss driven by radiation pressure from a hot spot located in the stream impact region on the stellar surface or accretion disc edge. The mass loss extracts angular momentum from the system, specifically  from the gainer.
The best model indicates that  \var\,already passed by a non-conservative
epoch 31400 years ago, and now it is in a conservative regime. The fact that observations actually show some mass loss suggests a quasi-conservative stage instead, as stated previously.

We find  two bright regions along the disc rim at longitudes $\lambda_{hs}$ = 325\dg and $\lambda_{bs}$ = 107\dg; the first one 1.21 times hotter than the disc rim, with 16\dg angular radius covering 9\% of the outer disc rim, and the second one 1.12 times hotter than the disc rim, with 54\dg angular radius, covering 30\% of the outer disc rim. These regions might be tentatively identified with shock regions revealed in hydrodynamical simulations of mass transfer in \var\,by Nazarenko \& Glazunova (2006) and in close binaries in general by Bisikalo et al. (1998, 1999, 2003).

Interestingly, the hot spot has also been inferred  from spectropolarimetric observations by Lomax et al. (2012).
They report a hot spot at the disc outer edge with a projected linear extension between 22 and 33 $R_{\odot}$,
equivalent to angular sizes 27.4 and 60.8 degrees (7.6 and 16.9 \% of the outer disc rim), depending on the photometric bandpass. The center of the hot spot is roughly at orbital phase 0.47, near the position of the hot spot found by us at orbital phase 0.40 ($\lambda_{hs}$ = 325\dg).

\subsection{Comparison with previous evolutionary studies}

We find \var\,in a Case-B mass transfer stage with an age of 2.30 $ \times$
10$^{7}$ yr.   The donor has exhausted hydrogen in its core and the gainer has a hydrogen concentration $X_{ch}$ = 0.43. The parameters for the best non-conservative evolutionary model  are given in Table 2. The late evolutionary stage of the donor is consistent with
the spectroscopic finding of He enrichment and extreme CNO cycling by Balachandran et al. (1986).


The evolutionary stage of \var\,was previously investigated by DGL94. These authors determined that a conservative solution led to several discrepancies with observed data. They find that \var\,is reaching the end of the mass transfer, having evolved in a non-conservative way from an initial system of 9 M$_{\odot}$ + 7.65 M$_{\odot}$ and orbital period 4 days (our best model starts with 12 M$_{\odot}$ + 7.2 M$_{\odot}$ and orbital period 2.5 days).
Their model parameters are given in Table 2 along with those of our best model.

In general, the stellar parameters derived from our evolutionary solution compares relatively well with those of DGL94. However, significant differences can be observed in  the initial masses and the initial orbital period. In addition, DGL94 model gives a 9 Myr older binary, a slightly more inflated donor and a mass transfer rate 75\% larger than ours.

The above differences are not negligible and point to some
physical difference between both models. One of the possible
reasons for the differences might be the more massive binary
considered in our solution, along with the shorter initial orbital
period. This means a faster evolution (a younger system as
observed)  but also that non-conservative effects should be
stronger than in the case studied by DGL94.  Another possible
reason is the way both models deal with mass loss, usually
parametrized with the parameter $\beta$, defined as the mass
fraction accreted by the gainer, so 1-$\beta$ is the mass fraction
lost by the system. DGL94 assume $\beta$ constant through the
whole mass transfer process whereas Rensbergen et al. consider
$\beta$ variable according to the physics of their model. In any
case, we caution that we cannot go further  interpreting the
differences between both models since the mass loss processes are
still not well understood, neither included with all physical
details in the models. At this stage, these differences can be
interpreted as upper estimates for the accuracy of the initial
binary parameters, its age and $\dot{M}$.


\begin{table*}
\centering
 \caption{The parameters of the Van Rensbergen et
al.  (2008, labeled vR08) models that best fit the $\beta$
Lyr data of Table 1 along with those of the model by De Greve
\& Linnell (1994, labeled DGL94).   Subindexes ``s'' and ``w'' refer to the strong and
weak interaction models for the same set of initial parameters, 
where the weak case is the best solution followed by the strong case, as discussed in the text.
The hydrogen and helium core
mass fractions  ($X_{c}$ and $Y_{c}$) are given for the cool and
hot star. Errors are given when available, corresponding approximately to
half of the grid step at a given parameter.}
 \begin{tabular}{@{}lrrrlrrr@{}}
 \hline
{\rm quantity}  &{\rm vR08$_{w}$} &{\rm vR08$_{s}$} &{\rm DGL94 } &  {\rm quantity} &{\rm vR08$_{w}$} &{\rm vR08$_{s}$} &{\rm DGL94}\\
 \hline
age (yr) & 2.30070(4)E7   &2.30457(1)E7 &3.1924E7   & period (d) &12.95(9) &12.95(8) &13.23  \\
$M_{c}$  (M$_{\odot}$) &2.248(6) &2.231(6) &2.95 & $M_{h}$ (M$_{\odot}$)  &13.308(6) &13.251(6) &13.27 \\
$\dot{M_{c}}$ (M$_{\odot}$ yr$^{-1}$ ) &-1.58(11)E-5   &-1.38(2)E-5 &-2.77E-05  &$\dot{M_{h}}$   (M$_{\odot}$ yr$^{-1}$) & 1.44(2)E-5  &1.47(4)E-5  &2.57E-5 \\
log $T_{c}$ ($K$)&4.162(1)   &4.161(1) &4.135  &log $T_{h}$ ( $K$) &4.482(1) &4.482(1) &4.444\\
log $L_{c}$ (L$_{\odot}$) &3.917(1) &3.911(1) &3.917 &log $L_{h}$ (L$_{\odot}$)&4.442(1) &4.449(1) &4.329\\
$R_{c}$ (R$_{\odot}$)&14.36(5)   &14.29(3) &16.26 &$R_{h}$ (R$_{\odot}$ ) &6.023(2) &6.076(2) &6.33   \\
$X_{cc}$ &0.00(0) &0.00(0)  &0.000 &$X_{ch}$ &0.43(0) &0.42(0) &0.479\\
$Y_{cc}$ & 0.98(0) &0.98(0) &0.980 &$Y_{ch}$ &0.55(0) &0.56(0) &0.501    \\
\hline
\end{tabular}
\end{table*}

\subsection{The future of \var}

According to the best synthetic model presented in Section 3,  the donor will lose contact with its Roche lobe after 72100 yr, becoming a detached binary, with stellar components of 1.78 M$_{\odot}$ and 13.77 M$_{\odot}$. The orbital period will be 23.47 days and the stars  will be in basically the same present evolutionary stage.
The model does not cover the posterior evolution of the gainer, but if the disc and rapid gainer rotation still persist at that time, the system might be easily classified like a binary Be star. This conclusion depends on how the angular momentum and circumstellar mass are distributed during the interaction epochs, something that future models
 need to consider carefully.



\begin{figure*}
\scalebox{1}[1]{\includegraphics[width=1.0\textwidth]{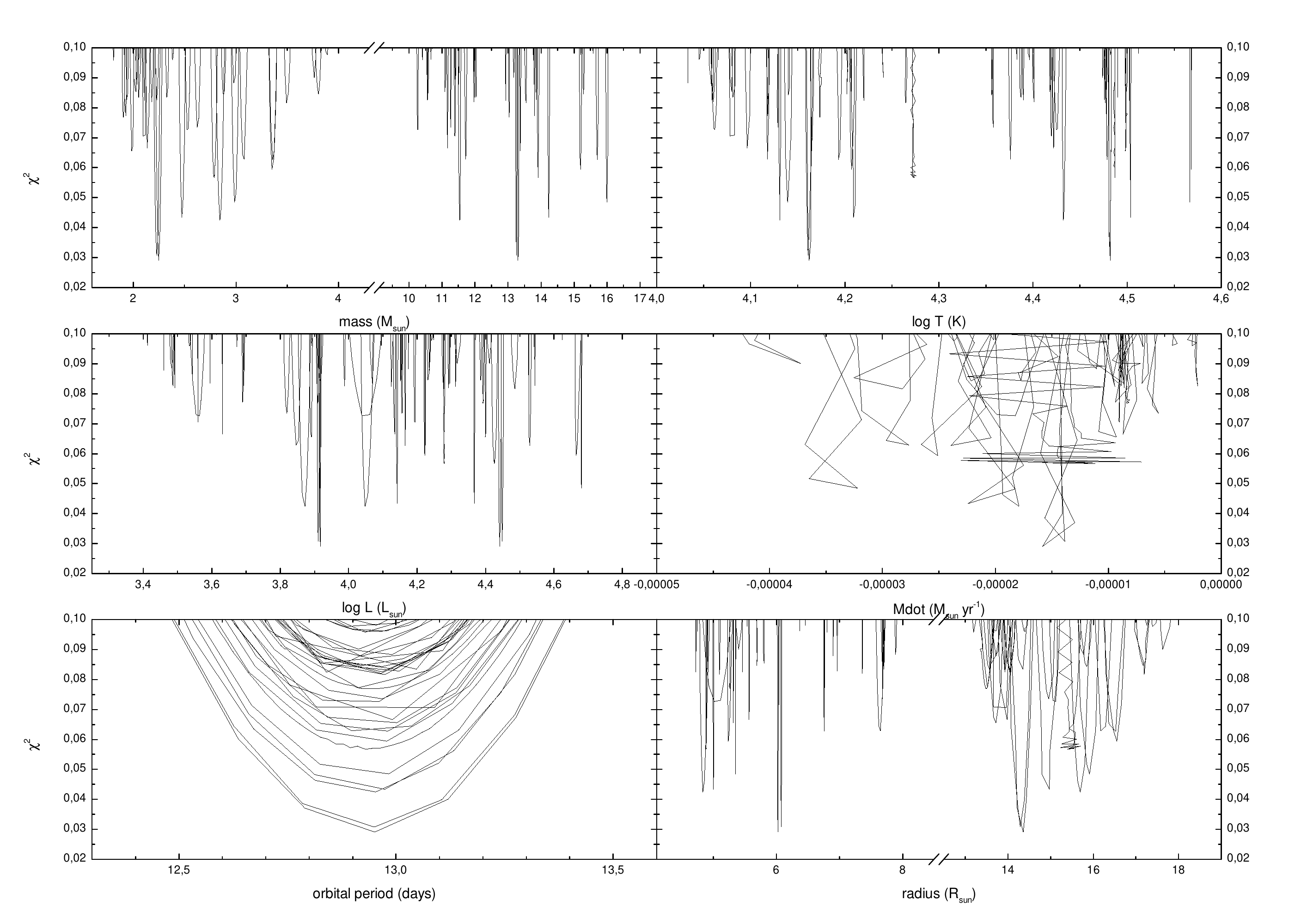}}
\caption{$\chi^{2}$ for all studied models versus selected parameters. Only the regions around the $\chi^{2}$ minima are shown. Temperatures, luminosities, masses and radii are given for {\em both components} in the corresponding panel. The significance of the best fit represented by the absolute minima, can be evaluated from the depth and proximity of secondary minima. }
  \label{x}
\end{figure*}

\section{Conclusions}

In this paper we fit the published optical light curve of \var\,with a light curve synthesis code based on the simplex algorithm, considering stellar plus disc components. In addition, the evolutionary stage of the binary was studied by comparing our best system parameters with those of grids of synthetic evolutionary tracks.  Hence, we have made use of the published data in an entirely new and complementary way. A comparison with earlier studies was also given. The main results of our research are:\\

\begin{itemize}
\item The $V$ light curve can be modeled with an intermediate mass semidetached binary including an accretion disc around the gainer.  The best system parameters including stellar and disc temperatures, masses, luminosities and sizes are given in Table 1.
\item Our study of the evolutionary stage of \var\,indicates that it is in a Case-B mass transfer stage, with age 2.30 $\times$ 10$^{7}$ years and $\dot{M} = 1.58\times 10^{-5}$ M$_{\odot}$ yr$^{-1}$. The donor has exhausted hydrogen in its core and the gainer is slightly evolved with central hydrogen fraction $X_{ch}$ = 0.43.
\item The best evolutionary model indicates that the system is found 147.000 years after 2nd Roche lobe contact. In this interaction episode the system has lost 0.03 M$_{\odot}$ into the interstellar medium and the gainer might have accreted  1.55  M$_{\odot}$. A fraction of this mass could still remain in the disc.
\item We provide arguments supporting the view that \var\,is in a quasi-conservative
evolutionary stage, having finished a relatively large mass loss episode 31400 years ago. By quasi-conservative, we mean a stage with a systemic mass loss much less that the mass transfer rate.
\item Discretization of the grid of synthetic models could explain the small discrepancy observed in donor temperature. At present we cannot quantify this effect, but argue it should be small. 
\item We find significant  evidence for two bright regions in the disc rim with temperatures  10 and 20\% higher than the disc outer edge temperature. They are located
at longitudes $\lambda$ = 107\dg and 325\dg respectively. The cooler bright spot region covers 30\% of the disc outer rim and the hotter one 9\%. These hot regions might be tentatively identified with those  shock regions revealed by hydrodynamical simulations of the mass transfer process.
\end{itemize}

\section{Acknowledgments}

 REM acknowledges support by Fondecyt grant 1110347 and the BASAL Centro de Astrofisica y Tecnologias Afines (CATA) PFB--06/2007. G.D.
acknowledges the financial support of the Ministry of Education and Science of the Republic of Serbia through the project 176004 ``Stellar physics''.
We acknowledge an anonymous referee for useful indications aimed to improve the first version of this manuscript.


\bsp
\label{lastpage}
\end{document}